\input{aipcheck}

\documentclass[
    ,final            
  ]
  {aipproc}

\usepackage{amsmath}

\layoutstyle{8x11single}

\newcommand{\Msun}{\mbox{$\mathrm{M}_{\odot}$}}

\begin{document}

\title{Chemically homogeneous evolution in massive binaries}


\author{S. E. de Mink$^{1,2}$}{
  address={$^{1}$Argelander Institut f\"ur Astronomie der Universit\"at Bonn University \\$^{2}$Astronomical Institute Utrecht, The Netherlands }
}

\author{M. Cantiello$^{1}$}{
}

\author{N. Langer$^{1,2}$}{
} 
\author{O. R. Pols$^{2}$}{
}

\begin{abstract}
Rotation can have severe consequences for the evolution of massive stars. It is now considered as one of the main parameters, alongside mass and metallicity that determine the final fate of single stars.  In massive, fast rotating stars mixing processes induced by rotation may be so efficient that helium produced in the center is mixed throughout the envelope.  Such stars evolve almost chemically homogeneously. At low metallicity they remain blue and compact, while they gradually evolve into Wolf-Rayet stars and possibly into progenitors of long gamma-ray bursts.  

In binaries this type of evolution may occur because of (I) tides in very close binaries, as a result of (II) spin up by mass transfer, as result of (III) a merger of the two stars and (IV) when one of the components in the binary was born with a very high initial rotation rate.  As these stars stay compact,  the evolutionary channels are very different from what classical binary evolutionary models predict.  
In this contribution we discuss examples of nearly chemically homogeneous evolution in very close tidally-locked binaries. Even in such very close massive binaries, the stars may remain compact and avoid mass transfer, while Roche lobe overflow and a merger would be inevitable in the classical picture.  This type of evolution may provide an alternative path to form tight Wolf-Rayet binaries and massive black hole binaries. 
\end{abstract}

\maketitle


\section{Introduction}

The rotation rate is considered as one of the main initial
stellar parameters, along with mass and metallicity, which determine
the evolution and fate of single stars. Rotation can deform the star, interplay with the
mass loss and trigger instabilities in the interior leading to
turbulent mixing in otherwise stable layers \citep[e.g.][]{Maeder+97}.
As rotationally induced mixing can bring processed material from the
core to the surface, it has been proposed as explanation for observed
surface abundance anomalies, such as a nitrogen enrichment found in
several massive main-sequence stars
\citep[e.g.][]{Walborn76,Maeder+Meynet+review00,Heger+Langer00}.
Rotation has also been successfully invoked to explain various phenomena  \citep[][and references therein]{MaederMeynet2000_Review}:  the ratio between O-stars and various types of Wolf-Rayet stars at different metallicities \citep{MeynetMaeder05_wolfrayetpopulations}, the production of nitrogen in the early universe \citep{MeynetMaeder02_primaryNitrogen},  the (apparent) presence of multiple populations in intermediate-age and globular clusters \citep{Decressin+07a, Decressin+07b, Bastian+DeMink09}, the variety of core collapse supernovae \citep{Georgy09} and the formation of progenitors of long gamma-ray bursts \citep{Yoon+Langer05}.

In models of rapidly rotating, massive stars, rotational mixing can
efficiently transport centrally produced helium throughout the stellar
envelope. Instead of expanding during core H-burning as non-rotating
models do, they stay compact, become more luminous and move blue-wards
in the Hertzsprung-Russell diagram \citep{Maeder87}. This type of
evolution is often referred to as (quasi-)chemically homogeneous
evolution and has been proposed as a pathway for the formation of long gamma-ray
burst progenitors \citep{Yoon+06,Woosley+06}.

High rotation rates can be achieved in binary systems due to
mass and angular momentum transfer \citep{Cantiello+07}, by tidal interaction in
close binaries \citep{Detmers+08} and, possibly, when the two stars in a binary merge.  In this contribution we discuss the possibility and consequences of chemically homogenous evolution in close massive binaries, which according to classical evolutionary models would inevitably experience Roche lobe overflow.



\section{Stellar evolution code}\label{sec:code}  

We model the evolution of rotating massive stars using the 1D
hydrodynamic stellar evolution code described by \citet{Yoon+06} and
\citet{Petrovic+05_GRB}, which includes the effects of rotation on the
stellar structure and the transport of angular momentum and chemical
species via rotationally induced hydrodynamic instabilities
\citep{Heger+00}.
We refer to \citep{DeMink+09} for a full description of the code.

\begin{figure}[t]
 \includegraphics[width=0.5\textwidth]{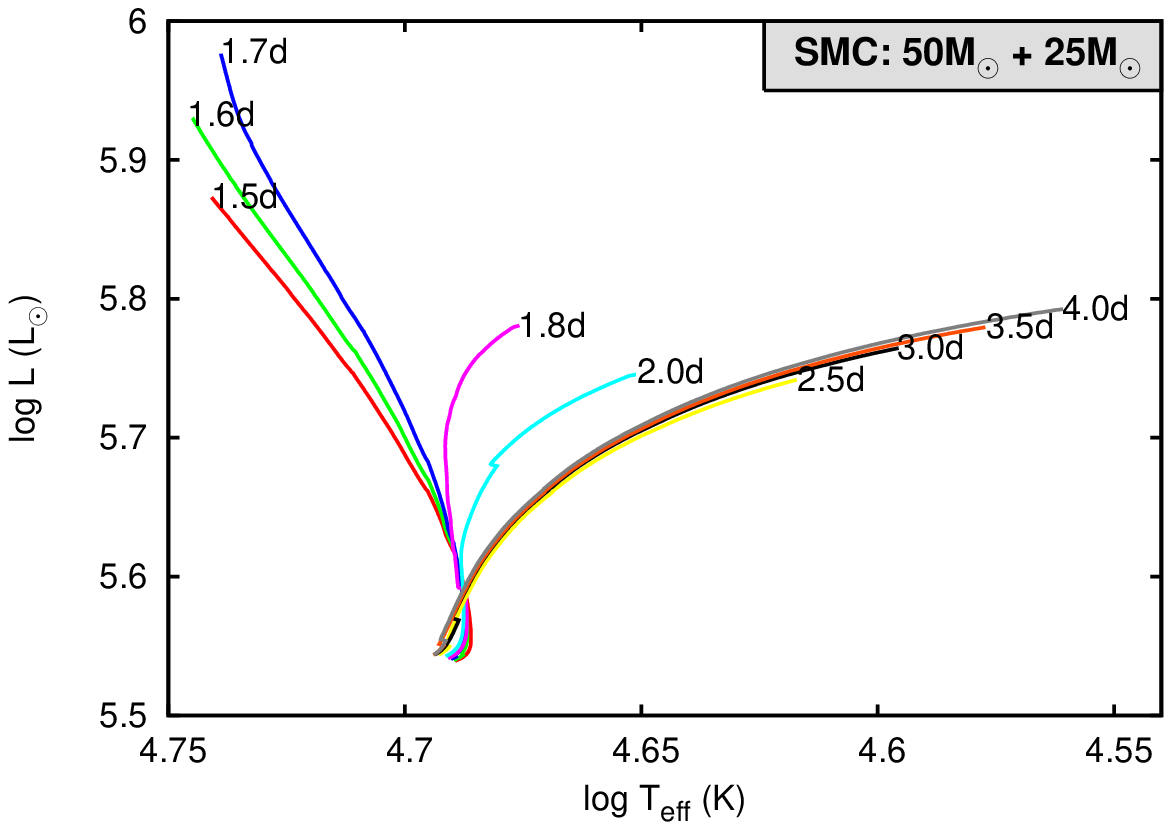}
  \includegraphics[width=0.5\textwidth]{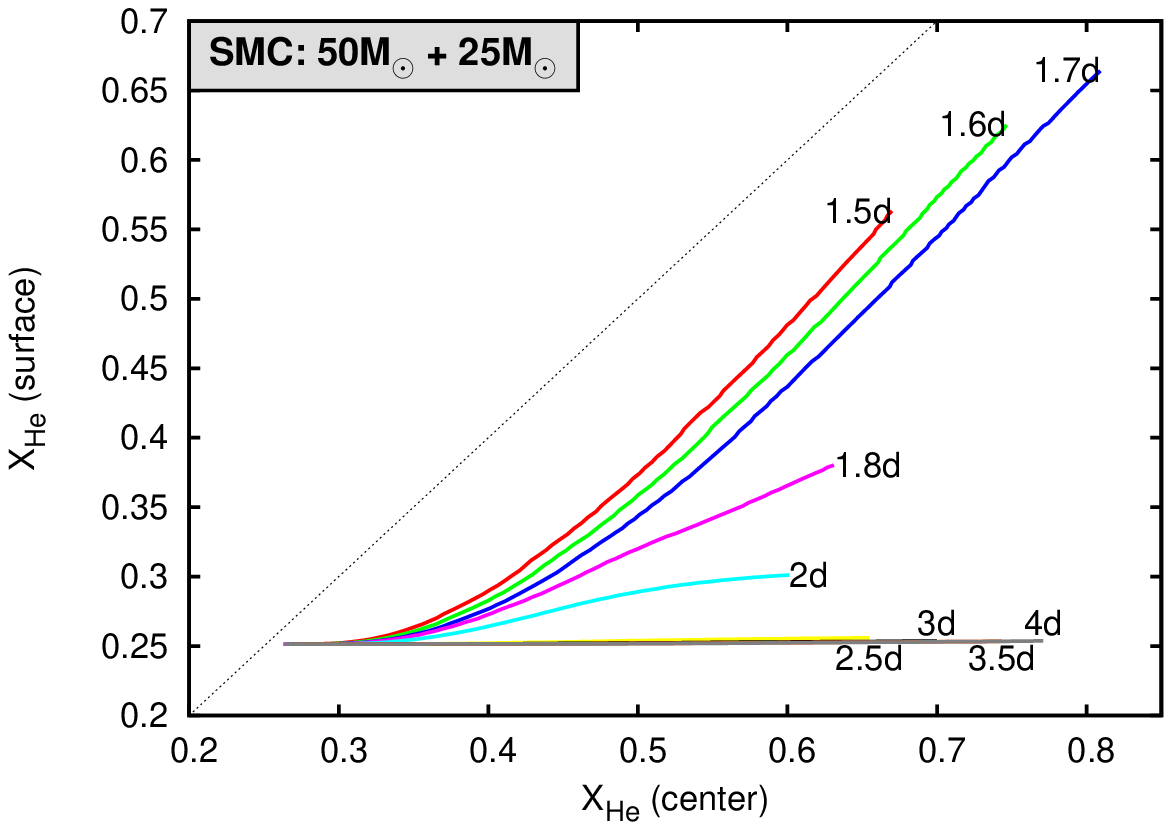}
  \caption{Left: The evolution from the onset of central H burning until the
    moment of Roche-lobe overflow for a 50\Msun~star in a binary
    with a 25\Msun~companion (not plotted) with initial orbital
    periods between 1.5 and 4 days. Right: Helium abundance at the surface as a function of the helium
  abundance in the center for the same systems.    
     }
  \label{lmc_hrd}
\end{figure}

\section {Chemically homogenous evolution in very massive binaries
  (50\Msun$+$25\Msun)} \label{bin:lmc} \label{sec:lmc}

In massive binaries, rotational mixing can be so efficient that
the change in the chemical profile leads to significant structural
changes.  In this section we discuss models for which we adopt a primary mass of 50\Msun, a
secondary mass of 25\Msun, orbital periods varying from 1.5 to 4
days, assuming an SMC composition.  
Although such massive close systems are rare, observational
counterparts do exist, for example two of the four massive binaries 
presented by \citet{Massey+02} which are located in the R136 cluster
at the center of the 30 Doradus nebula in the Large Magellanic Cloud:
R136-38\footnote{$M_1=56.9\pm0.6\Msun$, $M_2=23.4\pm0.2\Msun$ and $P_{\rm
orb} = 3.39$d} and R136-42\footnote{$M_1 =40.3\pm0.1\Msun $, $M_2
=32.6\pm0.1\Msun$ and $P_{\rm orb} =2.89$d}.
 Another example with an even closer orbit is [L72]~LH~54-425\footnote{
$M_1=47\pm2\Msun$, $M_2 =28\pm1\Msun$ and $P_{\rm orb} =2.25$d}
located in the LH~54~OB association in the Large Magellanic Cloud
\citep{Williams+08}. All three binary systems have O-type main-sequence
components, which reside well within their Roche lobes.

Figure~\ref{lmc_hrd} shows the evolution of our models in the
Hertzsprung-Russell diagram. The tracks are plotted until one of the stars in the binary
fills its Roche lobe  (not necessarily the primary, see
below). At the onset of hydrogen burning their location in the
diagram is very similar, although the stars in tighter binaries, which
rotate faster, are slightly cooler and bigger. This is a direct consequence of
the centrifugal accelaration.  As they evolve their tracks start to deviate.
The wider systems ($P_{\rm orb} > 2.0$d) evolve similarly to
non-rotating stars: they expand during core hydrogen burning, evolving
towards cooler temperatures until they fill their Roche lobe. Their
evolutionary tracks overlap in the Hertzsprung-Russell diagram.

The primaries in the tighter systems ($P_{\rm orb} < 2.0$d) behave very
differently: they evolve left- and upward in the HR-diagram, becoming
hotter and more luminous while they stay relatively compact.  The
transition in the morphology of the tracks around $P_{\rm orb} \approx
2$ days is analogous to the bifurcation found for fast rotating single
stars \citep{Maeder87,Yoon+Langer05}.

\subsubsection*{Surface abundances}

Rotational mixing is so efficient in these systems that even a large
amount of helium can be transported to the
surface. Figure~\ref{lmc_hrd} depicts the helium mass fraction at the
surface as a function of the helium mass fraction in the center.  In
the hypothetical case that mixing would be extremely efficient
throughout the whole star, the surface helium abundance would be equal
to the central helium abundance at all time. This is indicated by the
dotted line.  For the widest systems, the surface helium mass fraction
is not affected by rotational mixing, while for the tighter
systems $X_{\rm He}$ reaches up to 65\%. They follow the evolution of
chemically homogeneous stars.  Figure~\ref{lmc_hrd} also shows that for
each system, mass transfer starts before all hydrogen is converted
into helium in the center. Note that the highest central He mass
fraction when mass transfer starts is reached in the 1.7~day system
(more than 80\%), whereas on the basis of standard binary evolution
theory
\citep[e.g.][]{Kippenhahn+Weigert67} one would expect this to occur in
the widest system.  This anomalous behavior is connected to the
evolution of the radius, as discussed below.

\begin{figure}[t]
\centering
\begin{tabular}{cc}
\includegraphics[width=8.5cm]{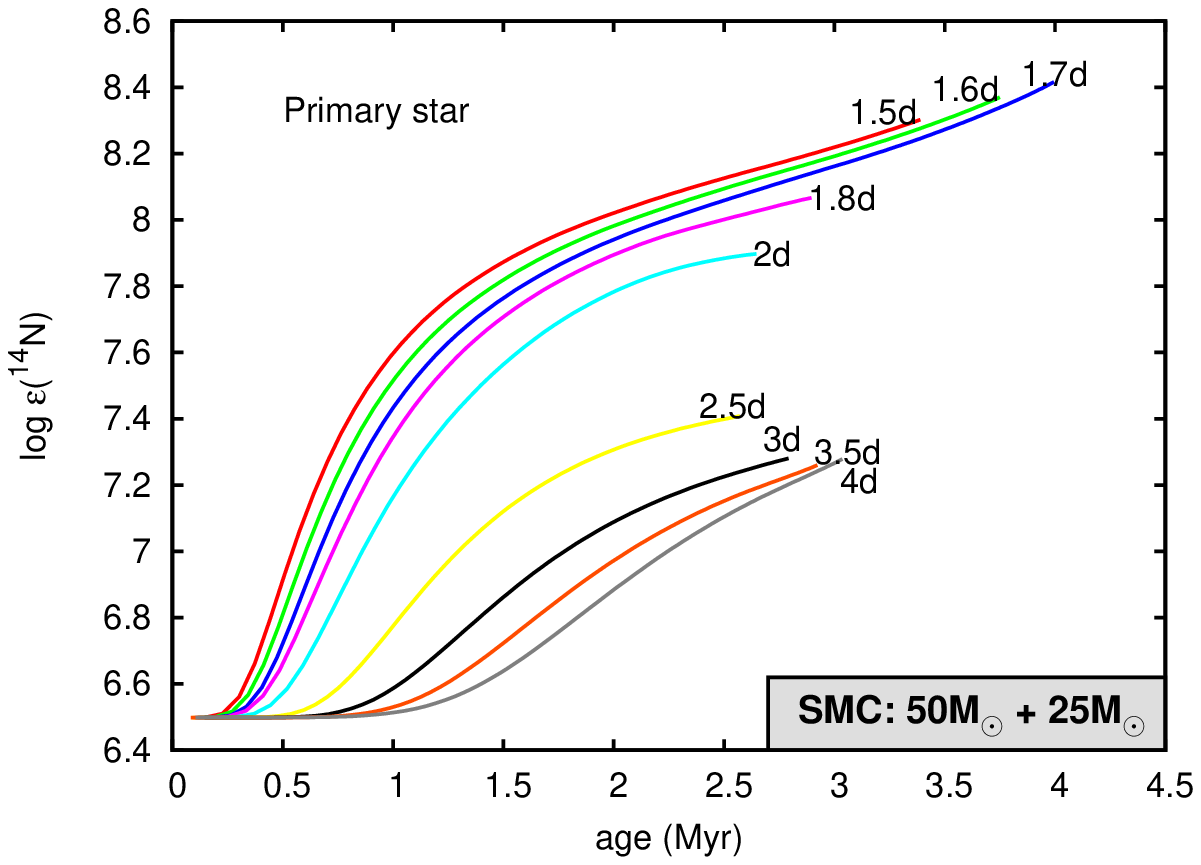} & \includegraphics[width=8.5cm]{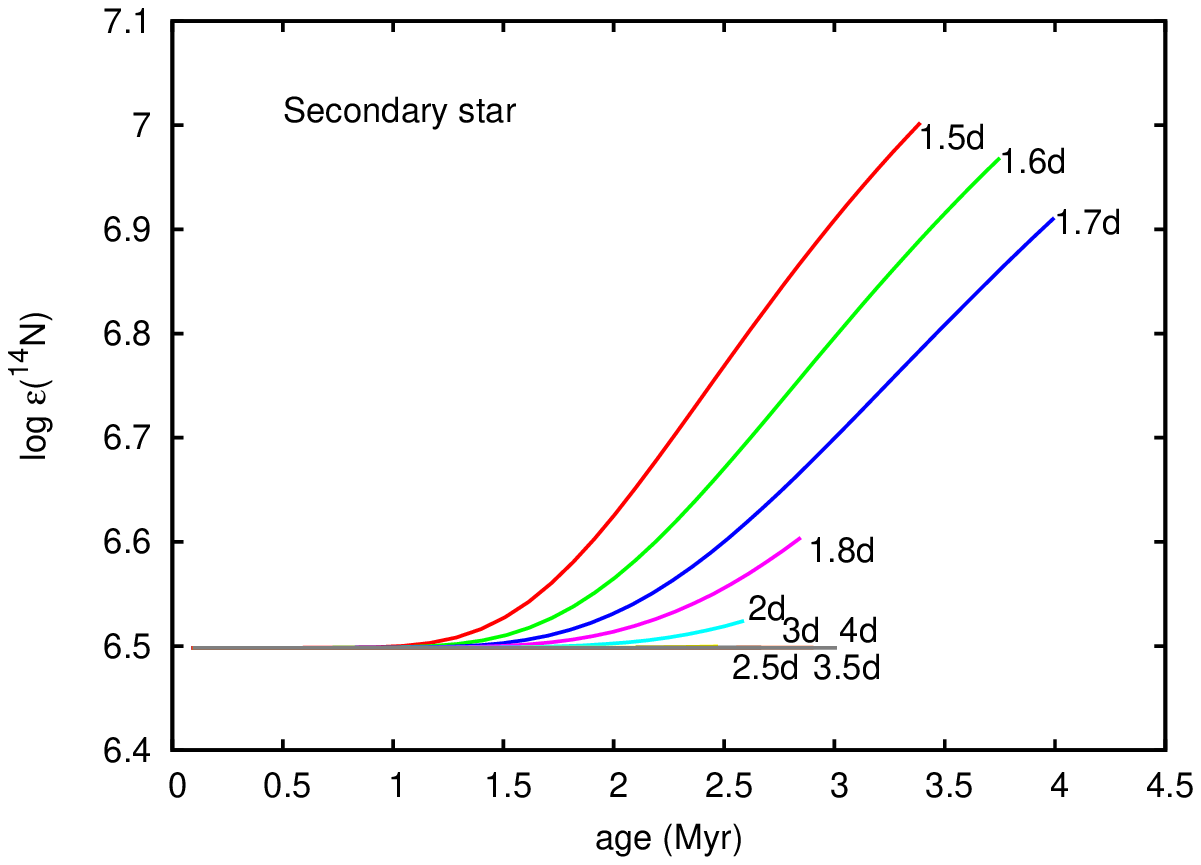}\\
\includegraphics[width=8.5cm]{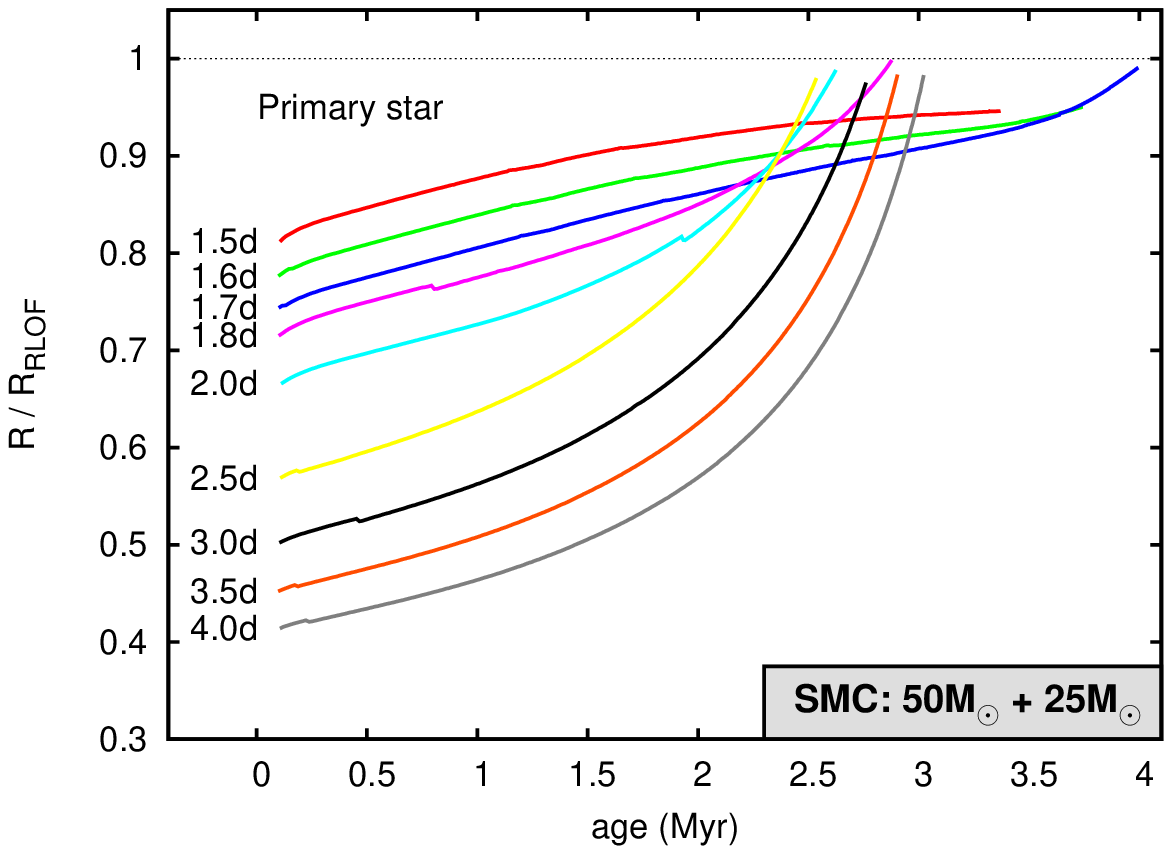}& \includegraphics[width=8.5cm]{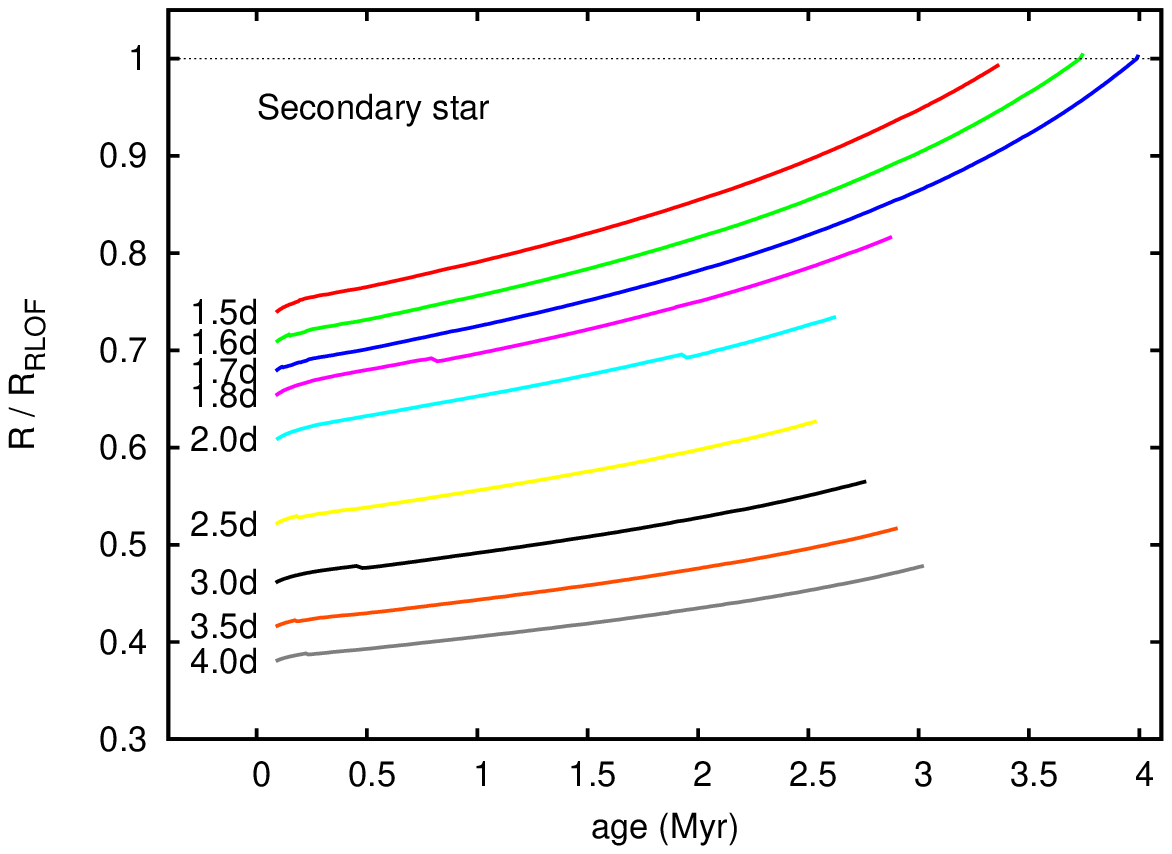}\\
\end{tabular}
\caption{{\bf First row:} Nitrogen abundance as a function of age at the
  surface for the primary and secondary star of the same systems as
  plotted in Figure~\ref{lmc_hrd}. Note the different scales.
  {\bf Second row:}  Radius as fraction of the Roche-lobe radius for the primary
  (left panel) and secondary star (right panel).
  } \label{lmc_abun}
\end{figure}

All systems show large enhancements of nitrogen at the surface of the
primary, see Fig.~\ref{lmc_abun}. The wider systems are enhanced by up
to 0.9 dex.  In the tight systems the enhancement reaches almost 2
dex. This extreme increase is partly due to the fact that abundance is
measured relative to hydrogen, which is significantly depleted at the
surface of the primaries in the tightest systems. In fact, due to rotational mixing, H is transported from the envelope to the core as well, where it is burned to He.
Also the secondary stars show nitrogen
surface enhancements, of up to 0.5 dex.

\subsubsection*{Evolution of the radius}

The increase of helium in the envelopes of the primary stars in the
tightest binaries leads to a decrease of the opacity and an increase
in mean molecular weight in the outer layers, resulting in more
luminous and more compact stars.  In Fig.~\ref{lmc_abun} we plot the
stellar radii as a fraction of their Roche-lobe radii.  The primary
stars in the wider systems expand and fill their Roche lobe after
about 2.5-3 Myr.  Contrary to what one might expect, we find that
Roche-lobe overflow is delayed in tighter binaries.  Whereas classical
binary evolution theory predicts that the primary star is the first to
fill its Roche lobe, we find instead that for systems with $P_{\rm
orb} \le 1.7$~days it is the less massive secondary star that
starts to transfer mass towards the primary. 
This is because the efficiency of rotational mixing increses with increasing stellar mass [e.g. 14], making the effect on the stellar structure more significant in the primary star.
During this phase of
reverse mass transfer from the less massive to the more massive star,
the orbit widens. Nevertheless we find that, if we continue our
calculations, the two stars come into contact shortly after the onset
of mass transfer and the stars are likely to merge.

\subsubsection*{Expected trends with system mass and mass ratio}

In more massive systems the effect of rotational mixing becomes
stronger in both components. This may allow for chemically
homogeneous evolution to occur in binary systems with wider orbits.
If the mass ratio is closer to one, $M_2 \approx M_1$, the effect of
rotational mixing becomes comparable in both stars, and they may both
evolve along an almost chemically homogeneous evolution track. In that
case both stars can stay within their Roche lobe and gradually become
two compact WR stars in a tight orbit.
In a system with a more extreme mass ratio, $M_2\ll M_1$, the
secondary star will hardly evolve or expand during the core
hydrogen-burning lifetime of the primary.  Also in this case
Roche-lobe overflow may be avoided during the core hydrogen-burning
lifetime of the primary component, leading to the formation of a
Wolf-Rayet star with a main-sequence companion in a tight orbit.

\section{Discussion} \label{sec:dis} 
We have shown that rotational mixing, if it is as efficient as assumed
in our models, can lead to chemically homogeneous evolution for
tight binaries with a 50\Msun~primary. In these models the primary
star stays so compact that the secondary star is the first to fill its
Roche lobe.

This peculiar behavior of the radius of stars, which are efficiently
mixed, has been noted in models of rapidly rotating massive single
stars \citep{Maeder87} and has been suggested as an evolutionary
channel for the progenitors of long gamma-ray bursts
\citep{Yoon+06, Woosley+06} in the collapsar scenario
\cite{Woosley93}.  In single stars this type of evolution only occurs
at low metallicity, because at solar metallicity mass and angular
momentum loss in the form of a stellar wind spins down the stars and
prevents initially rapidly rotating stars from evolving chemically
homogeneously \citep{Yoon+06, Brott+09}. In a close binary tides can
replenish the angular momentum, opening the possibility for chemically
homogeneous evolution in the solar neighborhood.

The binary models presented here all evolve into contact, but Roche-lobe overflow may be
avoided altogether in systems in which the secondary stays compact,
either because it also evolves chemically homogeneously, which may
occur if $M_1 \approx M_2$, or because it evolves on a much longer
timescale than the primary, when $M_2\ll M_1$. Whereas standard binary
evolution theory predicts that the shorter the orbital period, the
earlier mass transfer sets in, we find that binaries with the lowest
orbital periods may avoid the onset of mass transfer altogether.  This
evolutionary scenario does not fit in the traditional classification of
interacting binaries into Case~{\it A}, {\it B} and {\it C}, based on
the evolutionary stage of the primary component at the onset of mass
transfer \citep{Kippenhahn+Weigert67,Lauterborn70}.  In the remainder
of this contribution we will refer to this new case of binary evolution, in
which mass transfer is delayed or avoided altogether as a result of
very efficient internal mixing, as Case~{\it M}.

The massive and tight systems in which Case~{\it M} can occur are
rare. Additional mixing processes induced by the presence of the
companion star, which may be important in such systems, will widen the
parameter space in which Case~{\it M} can occur: it would lower the minimum
mass for the primary star and increase the orbital period below which
this type of evolution occurs.  The massive LMC binary
[L72]~LH~54-425, with an orbital period of 2.25~d
\citep{Williams+08} may be a candidate
for this type of evolution. Another interesting case is the galactic
binary WR20a, which consists of two core hydrogen burning stars of
$82.7\pm5.5\Msun$ and $81.9\pm5.5\Msun$ in an orbit of 3.69~d. Both stars
are so compact that they are detached. The surface abundances show
evidence of rotational mixing: a nitrogen abundance of six times
solar is observed and carbon is depleted \citep{Bonanos+04, Rauw+05}.

\subsubsection*{Short-period Wolf-Rayet and black-hole binaries  }

If Roche-lobe overflow is avoided throughout the core hydrogen-burning
phase of the primary star, both stars will stay compact while the
primary gradually becomes a helium star and can  be observed as a
Wolf-Rayet star.  Initially the Wolf-Rayet star will be more massive
than its main sequence companion, but mass loss due to the strong
stellar wind may reverse the mass ratio, especially in systems which
started with nearly equal masses. Examples of observed short-period
Wolf-Rayet binaries with a main-sequence companion are
CQ~Cep
, CX~Cep
, HD~193576 
and the very massive system HD~311884
\citep{vanderHucht01}. Such
systems are thought to be the result of very non-conservative mass
transfer or a common envelope phase
\citep[e.g.][]{Petrovic+05_WR}. Case~M is an alternative
formation scenario which does not involve mass transfer.

Case~{\it M} is also interesting in the light of massive black-hole
binaries. \citet{Orosz+07} recently published the stellar parameters
of M33~X-7. This system is located in the nearby galaxy Messier~33 and harbors one
of the most massive stellar black holes known to date, ${\rm
M_{bh}}=15.7\pm1.5\Msun$, orbiting a massive O star, ${\rm M_O}=70\pm
7\Msun$, which resides inside its Roche lobe in spite of the fact that
the orbit is very tight, $P_{\rm orb} = 3.45$~d.  The explanation for
the formation of this system with standard binary evolutionary models
involves a common-envelope phase that sets in after the end of core
helium burning (Case~{\it C}). This is because the progenitor of the black hole
must have had a radius much greater than the current orbital
separation.  This scenario is problematic as it requires that the
black-hole progenitor lost roughly ten times less mass before the
onset of Roche-lobe overflow than what is currently predicted by
stellar evolution models
\citep{Orosz+07}.  An additional problem is that the most likely
outcome of the common envelope phase would be a merger, as the
envelopes of massive stars are tightly bound
\citep{Podsiadlowski+03}.  However, see Valsecchi et al. these proceedings and \citep{Abubekerov+09}. 
In the Case~{\it M} scenario the black-hole
progenitor can stay compact and avoid Roche-lobe overflow, at least
until the end of core helium burning. This way the star retains its
envelope.  Whether this scenario can explain all the system parameters, remains to be investigated \citep[see also][and references therein]{Moreno-Mendez+08, Moreno-Mendez+09}.
%
%
%
Homogeneous evolution helps to explain
the high mass of the black hole, but the short orbital period poses a
difficulty also for Case~{\it M}. This is because strong mass loss during the Wolf-Rayet
life time will widen the orbit. 

The subsequent evolution of tight, rapidly rotating Wolf-Rayet binaries
remains to be investigated. If one or both members of the 
system can retain enough angular momentum to fulfill the collapsar
scenario \citep{Woosley93}, which may be hard as the tides can slow
down the stars \citep[e.g.][]{Detmers+08}, it may lead to the
production of one or even two long gamma-ray bursts.

\section{Conclusion} \label{sec:con}  

We investigated the effect of rotational mixing on the evolution of
detached short-period massive binaries using a state of the art
stellar evolution code.  
These systems often show eclipses and big radial velocity
variations, such that their stellar parameters, the rotation rate and
possibly their surface abundances can be determined with high
accuracy.  This enables a direct comparison between an observed system
and models computed with the appropriate stellar and binary
parameters.   Therefore we proposed in  \citet{DeMink+09} to use such systems as test cases for rotational mixing. An additional major advantage of using detached
main-sequence binaries is the constraint on the evolutionary
history. For a fast spinning apparently single star we do not know
whether it was born as a fast rotator or whether its rotation rate is
the result of mass transfer or merger event. In a detached
main-sequence binary we can exclude the occurrence of any mass transfer
phase since the onset of core-hydrogen burning.

In the most massive binaries we find that rotational instabilities can
efficiently mix centrally produced helium throughout the stellar
envelope of the primary. They follow the evolutionary path of chemically
homogeneous stars: they stay within their Roche lobe, being
over-luminous and blue compared to normal stars. Due to large amount
of nitrogen and helium at the surface these stars can be observed as
Wolf-Rayet stars with hydrogen in their spectra. In contrast to
standard binary evolution, we find that it is the less massive star in
these systems that fills its Roche lobe first.

There may be regions in the binary parameter space in which Roche-lobe
overflow can be avoided completely during the core hydrogen-burning
phase of the primary. The parameter space for this new evolutionary
scheme, which we denote Case~{\it M} to emphasize the important role of
mixing, increases if additional mixing processes play a role in such
massive systems.  It may provide an alternative channel for the
formation of tight Wolf-Rayet binaries with a main-sequence companion,
without the need for a mass transfer and common envelope phase to
bring the stars close together.  This scenario is also potentially
interesting for tight massive black hole binaries, such as the intriguing system M33~X-7
\citep{Orosz+07}.


\begin{theacknowledgments} We acknowledge the stellar evolution groups in Utrecht and Bonn for stimulating discussions and the LKBF for financial support. 
\end{theacknowledgments}



\bibliographystyle{aipproc}   

\bibliography{references}

\IfFileExists{\jobname.bbl}{}
 {\typeout{}
  \typeout{******************************************}
  \typeout{** Please run "bibtex \jobname" to optain}
  \typeout{** the bibliography and then re-run LaTeX}
  \typeout{** twice to fix the references!}
  \typeout{******************************************}
  \typeout{}
 }

\end{document}